\documentclass[10pt,aps,prx,twocolumn,superscriptaddress,floatfix]{revtex4-2}
\usepackage{graphicx}
\usepackage{xcolor}
\usepackage{amsmath}
\usepackage{amssymb}
\usepackage[colorlinks=true,linkcolor=blue,citecolor=blue,urlcolor=blue]{hyperref}
\begin{document}
\title{A molecular perspective on coordination, screening, and emergent length scales in lithium electrolytes}
\author{Amaury Coste}
\affiliation{Univ. Grenoble Alpes, CEA, IRIG-MEM-LSim, 38054 Grenoble, France}
\author{Eva Zunzunegui-Bru}
\affiliation{Univ. Grenoble Alpes, Univ. Savoie Mont Blanc, CNRS, Grenoble INP, LEPMI, Grenoble, France}
\affiliation{Univ. Grenoble Alpes, CEA, IRIG-MEM-LSim, 38054 Grenoble, France}
\author{Ambroise van Roekeghem}
\affiliation{Univ. Grenoble Alpes, CEA, LITEN, 38054 Grenoble, France}
\author{Ioannis Skarmoutsos}
\affiliation{Laboratory of Physical Chemistry, Department of Chemistry, University of Ioannina, 45110 Ioannina, Greece}
\author{Stefano Mossa}
\affiliation{Univ. Grenoble Alpes, CEA, IRIG-MEM-LSim, 38054 Grenoble, France}
\email{stefano.mossa@cea.fr}
\date{\today}
\begin{abstract}
Lithium electrolytes are commonly described using separate conceptual frameworks for local coordination chemistry, electrostatic screening, and ionic transport. This separation is effective in dilute conditions but breaks down at higher concentration, where coordination, ion pairing, clustering, and collective dynamics become intrinsically coupled. In this Perspective, we develop a unified multiscale framework that links local coordination motifs, mesoscopic ionic organization, and macroscopic transport within a single physical picture. Guided primarily by representative examples from molecular simulations, spanning carbonate liquids, polymer electrolytes, concentrated systems, and confinement, we show that increasing concentration drives a systematic evolution from solvent-dominated Li$^+$ coordination to ion pairing, clustering, and correlated ionic domains. In this regime, screening and transport are not independent phenomena but emerge from the same underlying correlated structures. The resulting hierarchy provides both a conceptual framework for interpreting electrolyte behavior and a foundation for multiscale computational approaches, including machine-learning-assisted coarse-graining, that connect atomistic chemistry to macroscopic electrochemical performance. This Perspective implies that rational electrolyte design must simultaneously control short-range coordination, mesoscale organization, and collective electrostatic response.
\end{abstract}
\maketitle
\begin{figure*}[t]
\centering
\includegraphics[width=0.9\textwidth]{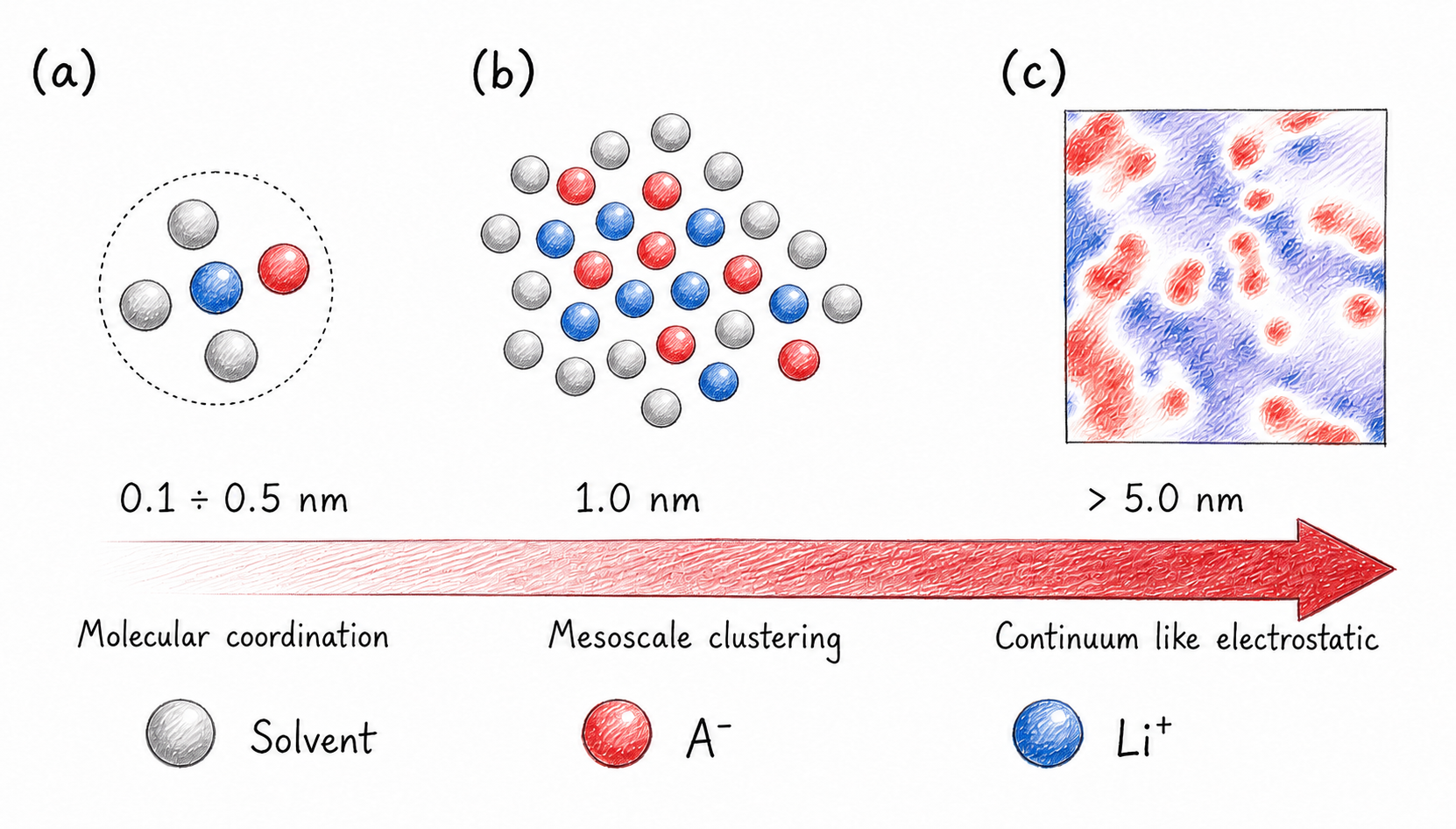}
\caption{\textbf{Hierarchy of structural and electrostatic correlations in liquid electrolytes.} Schematic illustration of the progressive coarse-graining from local coordination to collective electrostatics with increasing length scale. \textbf{(a) Coordination motif ($\sim0.1\!-\!0.5$~nm).} A lithium ion (blue) is coordinated by one counter-ion (red) and three solvent molecules (gray), defining an anisotropic local coordination motif. The dashed circle of radius $r$ delimits the first coordination shell, which sets the shortest structural length scale governing local ion--environment interactions. \textbf{(b) Correlated ionic clusters ($\sim1$~nm).} At intermediate scales, coordination motifs assemble into dynamically fluctuating, quasi-neutral ionic aggregates composed of contact and solvent-shared ion pairs together with larger clusters embedded in the solvent. This regime reflects the competition between short-range coordination and longer-range Coulomb interactions, leading to heterogeneous yet locally charge-balanced structures. \textbf{(c) Collective electrostatic response ($>5$~nm).} At larger scales, the discrete molecular structure is progressively coarse-grained into an effective medium described by collective charge-density fluctuations and screened electrostatic fields. Many-body correlations govern the long-wavelength electrostatic response. The bottom axis summarizes the corresponding hierarchy of length scales, from coordination motifs to correlated ionic clusters and finally to collective electrostatic behavior. Colors denote lithium ions (blue), anions (red), and solvent molecules (gray).}
\label{fig:coordination}
\end{figure*}
\section{Introduction}
Electrolytes remain a central bottleneck in electrochemical energy storage, as they simultaneously control ionic conductivity, interfacial chemistry, electrochemical stability, and power delivery~\cite{Xu2004,Xu2014}. This is especially critical in modern battery technologies~\cite{armand2008,goodenough2010}. For lithium-based systems, the most natural microscopic starting point is the local coordination environment of Li$^+$: the cation binds solvent molecules, polymer segments, or anions and forms a {\em dressed}~\cite{kjellander2018} object embedded in a fluctuating charged medium. This viewpoint is chemically transparent and has been extremely productive, especially for dilute liquid electrolytes.

That picture is nevertheless incomplete once one moves away from the dilute limit. Concentrated electrolytes, solvent-in-salt systems, polymer electrolytes, ionic-liquid-based formulations, and hybrid materials all exhibit strong correlations, pronounced deviations from ideal transport, and screening behavior that cannot be reduced to isolated solvation shells~\cite{borodin2009,borodin2016,feng2019,francelanord2019,skarmoutsos2025}. As the salt concentration increases, anions enter the first coordination shell, ion pairing and clustering become important, and extended ionic domains may emerge. The relevant physical objects are then no longer adequately described as bare ions, or even merely solvated ions, but as correlated ionic structures, whose identity depends on both local chemistry and collective organization.

The objective of this perspective is to develop a unified multiscale framework that connects these levels of description. The conceptual development of this framework is illustrated in Fig.~\ref{fig:coordination}. The central idea is that the physically relevant object evolves systematically with the observation scale, from local coordination motifs to correlated ionic domains. Accordingly, coordination, screening, and transport should not be viewed as separate phenomena, but as different manifestations of the same hierarchy of correlated ionic structures. Local coordination determines the elementary motifs from which ion pairs, clusters, and larger ionic domains are built. Those larger structures, in turn, govern dielectric response, screening, and the effective carriers relevant for charge transport.

The discussion is organized around representative examples, largely drawn from our own work where appropriate. Accordingly, the manuscript is intended as a conceptual perspective rather than as an exhaustive review of the extensive body of work in the field. We also emphasize, however, that the perspective developed here builds upon concepts and insights that are already present in the literature, albeit often in a fragmented or dispersed form. Where appropriate, we also highlight complementary theoretical developments that have contributed to establishing links between solvation, ionic association, electrostatic response, and transport.

Most of the results we consider originate from molecular simulations, which provide a consistent and controlled framework for analysis. This emphasis reflects the fact that experimental probes, while essential, are often not sufficiently selective to unambiguously isolate the most relevant microscopic degrees of freedom and their correlations. By contrast, simulations allow one to resolve shell composition, cluster statistics, and transport correlations at a level of detail that helps disentangle the underlying physical mechanisms, while the resulting physical picture is intended to remain broader than any single methodology. Beyond providing microscopic insight, these simulations naturally motivate a multiscale computational framework in which information is transferred from atomistic coordination chemistry to continuum electrochemical response.

We begin by examining local coordination in molecular and polymer electrolytes, and then introduce a hierarchy of length scales that links short-range structure to mesoscopic organization and electrostatic screening. We next address how confinement perturbs this hierarchy, before showing how the same framework naturally extends to transport, multiscale computational modeling, and electrolyte design. If established, such a framework could guide the rational design of electrolyte materials by explicitly connecting molecular-level structure to macroscopic transport properties. Within this picture, local coordination defines the effective charged entities, their aggregation governs screening, and the resulting correlated structures control transport.

Throughout this perspective, we use the terms ``motif'', ``cluster'', and ``domain'' in an operational sense. Specifically, ``motif'' denotes a local coordination environment surrounding one or several ions, ``cluster'' designates a dynamically associated ionic aggregate extending beyond the first coordination shell, and ``domain'' refers to larger-scale correlated regions that emerge at high concentration. Although the boundaries between these objects are not always sharply defined, this terminology provides a useful conceptual basis for discussing the progressive coarse-graining from molecular coordination to collective electrolyte behavior.
\begin{figure}[t]
\centering
\includegraphics[width=0.48\textwidth]{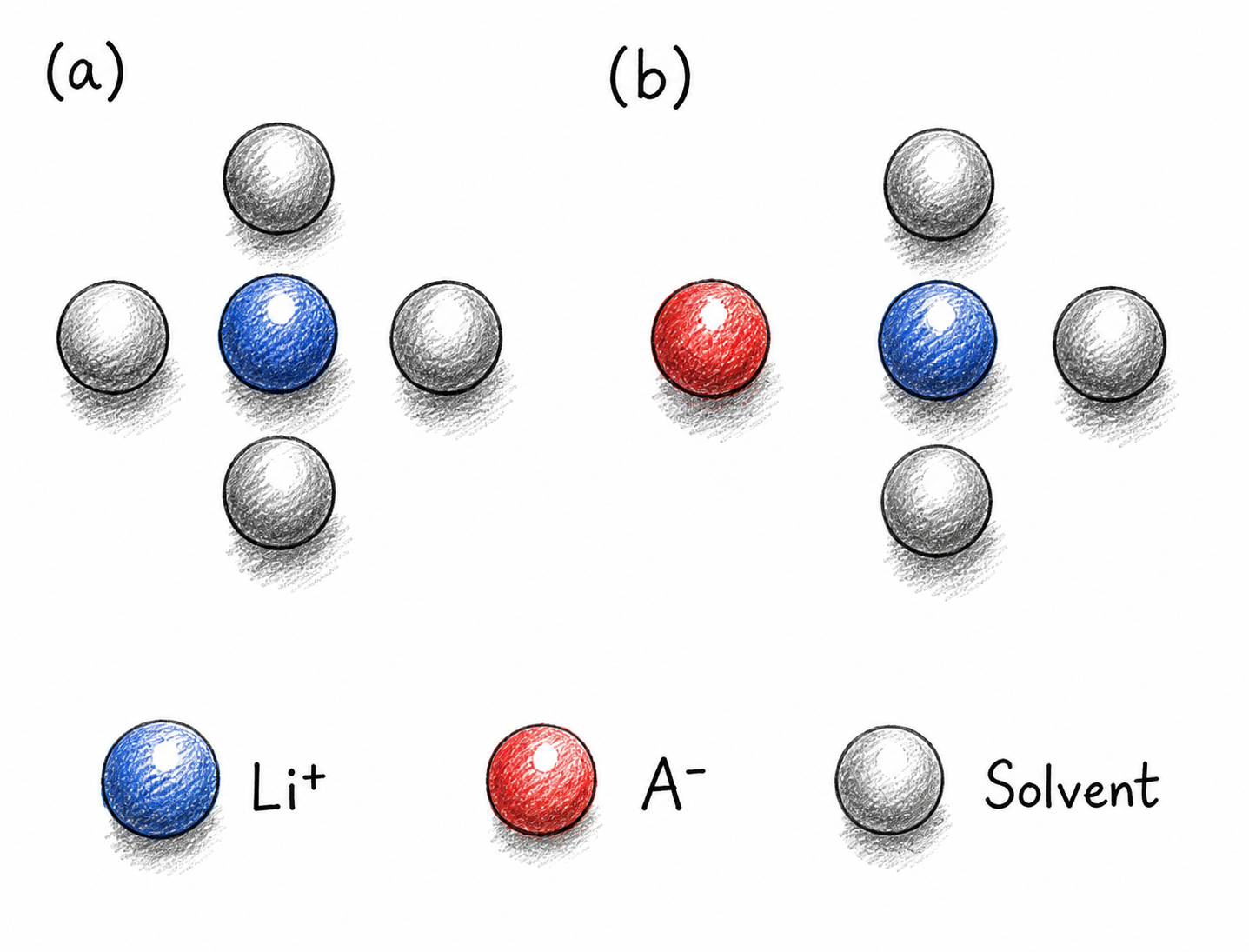}
\caption{\textbf{Evolution of the lithium coordination motif from dilute to concentrated electrolytes.} \textbf{(a) Dilute regime.} Li$^+$ is primarily coordinated by solvent molecules, forming a first coordination shell containing approximately four solvent ligands arranged in a nearly tetrahedral geometry. The effective lithium charge is only weakly screened, and the local structure is dominated by ion--solvent interactions. \textbf{(b) Concentrated regime.} As the salt concentration increases, anions progressively enter the first coordination shell, replacing solvent molecules and forming contact or solvent-separated ion pairs. The solvent coordination number decreases, while the dressed lithium ion becomes partially neutralized. This evolution transforms the elementary coordination motif from a solvent-dominated tetrahedral shell into a mixed cation--solvent--anion environment, providing the microscopic origin of ion pairing, cluster formation, and ultimately the correlated electrostatic behavior characteristic of concentrated electrolytes~\cite{jiang2016,ponnuchamy2018}.}
\label{fig:dilute_concentrated}
\end{figure}
\section{Local coordination as the elementary structural problem}
Local coordination provides the natural microscopic starting point for understanding electrolyte structure. At this scale, the central questions concern the identity of the species coordinating Li$^+$, the geometry of the first coordination shell, and the manner in which this local environment evolves with electrolyte composition and concentration. As discussed below, this local description already contains the seeds of the larger-scale organization that ultimately governs screening and transport.
\subsection{Carbonate solvents and the tetrahedral motif}
For conventional carbonate electrolytes, the classical starting point is the interaction of Li$^+$ with carbonyl oxygen atoms. Ethylene- (EC), propylene- (PC), and dimethyl-(DMC) carbonates differ significantly in molecular shape, dielectric character, and flexibility, yet they all offer strong donor sites to Li$^+$. A broad body of simulation and quantum-chemical work indicates that, at low concentration, Li$^+$ typically adopts a fourfold first-shell molecular coordination in these media, with approximate {\em tetrahedral} local order~\cite{soetens1998,li1999,masia2004,bhatt2010,bhatt2012,borodin2009,skarmoutsos2015}, as illustrated in Fig.~\ref{fig:dilute_concentrated}. This picture is chemically natural: the small cation favors strong short-range interactions with a limited number of donor oxygens, and fourfold coordination is consistent with both steric and electrostatic constraints.

At the same time, the tetrahedral motif should not be over-interpreted as a rigid molecular complex. The simulations of~\cite{skarmoutsos2015} make an important point: even in dilute carbonate mixtures, the local environment of Li$^+$ is embedded in substantial structural heterogeneity extending to distances more than five times larger than the first coordination-shell radius. This is already a first sign that the electrolyte cannot be reduced to a collection of isolated clusters. The shell is local, but its composition and dipolar environment are continuously influenced by larger-scale structural fluctuations. Moreover, the local tetrahedral arrangement should be understood as a preferred instantaneous coordination motif rather than as a unique or static molecular structure.

This point matters because the literature on carbonate solvation has long been marked by apparently conflicting coordination numbers extracted from Nuclear Magnetic Resonance, Raman spectroscopy, mass spectrometry, and simulations~\cite{hyodo1989,ogara1991,fukushima2001,bogle2013,yang2010,yuan2014}. Part of the discrepancy is methodological, but part is physical: the ``coordination number'' is not always a sharply defined observable. It depends on concentration, on the operational cutoff used, on whether one counts coordinating molecules or donor atoms, and on whether anions participate in the same shell. This distinction becomes particularly important at high salt concentration, where the molecular coordination number decreases because anions replace solvent molecules, while the donor-atom coordination may remain close to four, owing to bidentate anion coordination. What survives across methods is the robust importance of a short-range carbonyl-oxygen cage around Li$^+$, and the remarkable sensitivity of this cage to composition and concentration.
\subsection{Preferential solvation in mixed solvents}
Mixed solvents provide a clear demonstration that local coordination is not controlled by polarity alone. If electrostatics were the only criterion, the most polar cyclic carbonates would always dominate the first shell. Yet both simulation and electronic-structure work show a more subtle picture. For EC/DMC and EC/DMC/PC mixtures, DFT calculations on isolated clusters tend to favor more strongly polar components, whereas molecular dynamics in the liquid may predict enhanced DMC participation in the first coordination shell~\cite{borodin2009,borodin2016,skarmoutsos2015,ponnuchamy2018,bogle2013,yuan2014}. This is not a contradiction; it reflects the fact that local structure in the liquid is shaped by \emph{packing}, \emph{collective dipolar cancellation}, and the structural (compositional) \emph{fluctuations} inherent to the condensed phase, not only by isolated binding energies.

The simulations of~\cite{skarmoutsos2015} are especially illuminating on this point. Within a locally tetrahedral shell, the vector sum of the solvent dipoles can be strongly reduced when similar molecules occupy the shell in symmetric arrangements. Under such circumstances, the total shell dipole may actually increase when a less polar linear carbonate contributes, because it disrupts otherwise more complete dipole cancellation. The resulting preference for DMC in the liquid therefore does not imply stronger pairwise Li$^+$--DMC interactions, but rather reflects the optimization of the entire local coordination environment within the surrounding liquid structure. This observation carries an important implication for electrolyte design. The question is not simply which solvent binds Li$^+$ more strongly, but which mixture produces the most favorable balance of local geometry, shell dipole, and bulk transport.
\subsection{Anion entry into the first shell}
The most direct route by which local coordination becomes a problem of charged-fluid physics is through anion participation. In concentrated carbonate electrolytes, anions such as BF$_4^-$, PF$_6^-$, and ClO$_4^-$ need not remain outside the Li$^+$ solvation shell. Instead, they may enter the latter and form contact ion pairs or larger aggregates~\cite{kameda2007,cresce2015anion,chapman2017,jiang2016,ponnuchamy2018}, see Fig.~\ref{fig:dilute_concentrated}. This transition fundamentally alters the nature of the local coordination problem. A key result in this area is the combined femtosecond vibrational spectroscopy and DFT study of~\cite{jiang2016}, where it was found that in EC solutions above 0.5~M the average number of directly bound EC molecules falls to about two. Importantly, this reduction refers to the number of coordinating solvent molecules rather than to the total number of donor atoms surrounding Li$^+$. The qualitative implication is profound. The standard four-solvent tetrahedral picture remains a useful reference for dilute conditions, but it ceases to describe the dominant local coordination motif under many battery-relevant concentrations. The physically relevant object therefore becomes a correlated cation--solvent--anion complex rather than an isolated solvated ion.

This result was generalized and refined in the quantum-chemical study of~\cite{ponnuchamy2018}, where the most stable contact-ion-pair motifs in PC- and DMC-based electrolytes were found to involve roughly two solvent molecules and one anion around Li$^+$. Crucially, although the total coordination in terms of \emph{molecular} neighbors drops, the local donor-atom geometry can still preserve a quasi-tetrahedral arrangement when the anion coordinates in a bidentate manner. That is, local tetrahedrality survives in an \emph{interatomic} sense even as the molecular interpretation changes. This distinction reconciles the apparently conflicting coordination numbers reported in different experimental and computational studies, and illustrates that the physically relevant coordination descriptor depends on the specific question addressed. More generally, it illustrates that the appropriate coordination descriptor depends both on the level of structural description and on the physical property under consideration.

The transport implications follow naturally from this structural evolution, as discussed in more detail below. When an anion occupies the coordination shell, the dressed lithium object is less positively charged and often less mobile under an applied field. This reduced mobility arises from the combined effects of stronger ion association, the increased effective size of the correlated complex, and the need for cooperative rearrangements during transport. This provides a clean microscopic rationale for the conductivity loss at high salt concentration~\cite{jiang2016,ponnuchamy2018,kameda2007,cresce2015anion,chapman2017}. The point is not merely that there are ``more interactions'' at high concentration; rather, the identity of the charge carrier itself changes. The moving entity is no longer Li$^+$ plus a neutral shell, but a partially neutralized correlated complex.

The examples discussed above illustrate a general principle that extends well beyond carbonate electrolytes: local coordination defines the elementary structural motifs from which larger correlated ionic structures ultimately emerge. This observation provides the conceptual bridge to polymer electrolytes, where the same coordination principles operate within a much slower and more constrained dynamical environment.
\section{Polymer coordination and the extension of the same ideas to solids}
Polymer electrolytes are frequently discussed as a separate class of electrolytes, but many of the structural ideas developed above remain applicable. The underlying coordination principles are largely the same, although their consequences become inherently dynamical because ion transport is now coupled to the slow relaxation of the host matrix~\cite{berthier1983,borodin2006,coste2026}. The major difference is that the donor atoms belong to a deformable, slowly relaxing matrix rather than to a molecular solvent. As a result, coordination and transport become strongly entangled with the internal dynamics of the host.

In poly(ethylene oxide)-based electrolytes, Li$^+$ is stabilized by ether oxygens, and its migration is tied to polymer segmental motion~\cite{berthier1983,borodin2006}. In carbonate-based polymer electrolytes, such as LiPF$_6$/polypropylene carbonate, the donor groups are more directly reminiscent of liquid carbonate solvents, but the dynamical landscape is much slower. It is also characterized by pronounced dynamic heterogeneity, reflecting broad distributions of polymer relaxation times and coordination lifetimes. The recent simulations of~\cite{coste2026} show that such systems exhibit strong ionic correlations, only a limited fraction of truly free ions, and a large population of negatively charged clusters at the highest salt concentrations. Even when the local Li$^+$--carbonate interaction is not particularly strong on a pairwise basis, the network of ionic and polymeric constraints can suppress the mobility of the cation more than that of the anion at experimentally relevant temperatures.

This, again, reinforces the same message: local coordination defines the elementary structural motifs, but it does not by itself determine macroscopic transport. The polymer matrix modifies the statistics of site availability, residence times, and exchange pathways. What appears as a simple coordination shell in a static structural snapshot is, dynamically, a transient and constrained network. The lifetime and connectivity of this network become as important as its instantaneous structure. That network determines whether Li$^+$ moves by carrying a coordination shell, by exchanging among neighboring coordination sites, or by participating in the motion of larger ionic aggregates. Polymer electrolytes therefore reinforce, rather than contradict, the multiscale picture developed for liquid electrolytes.

Hybrid polymer/ceramic systems offer another instructive example. The recent analysis of SiO$_2$ nanoparticles in PEO/LiTFSI shows that adding passive nanoparticles can reduce, rather than enhance, ionic conductivity above the melting temperature of the polymer~\cite{dalmas2024}. The reduction in conductivity arises not only from geometric constraints but also, and more fundamentally, from a reorganization of local coordination and species distributions into distinct regimes separated by a concentration threshold. In other words, filler addition changes the phase space of accessible local coordination environments and, therefore, the statistics of transport pathways. The fact that the resulting behavior can be interpreted using the same coordination framework developed for liquid electrolytes highlights the unifying character of this description.

Finally, liquid-crystalline solid electrolytes provide an intermediate case in which ion-conducting domains become structurally ordered on nanometer length scales while still allowing local Li$^+$ hopping~\cite{khan2024}. These systems make explicit what is otherwise hidden in conventional electrolytes: local coordination motifs can self-organize into mesoscopic ion-conducting channels, and transport then depends on the architecture of the coordinated domains as much as on the local coordination shell itself.

Taken together, these examples demonstrate that polymer and solid electrolytes do not require a fundamentally different conceptual framework from liquid electrolytes. Rather, they extend the same hierarchy of coordination motifs, ionic aggregates, and correlated structures into a regime where the slow dynamics of the host matrix becomes an integral part of the transport process.
\begin{figure*}[t]
\centering
\includegraphics[width=0.95\textwidth]{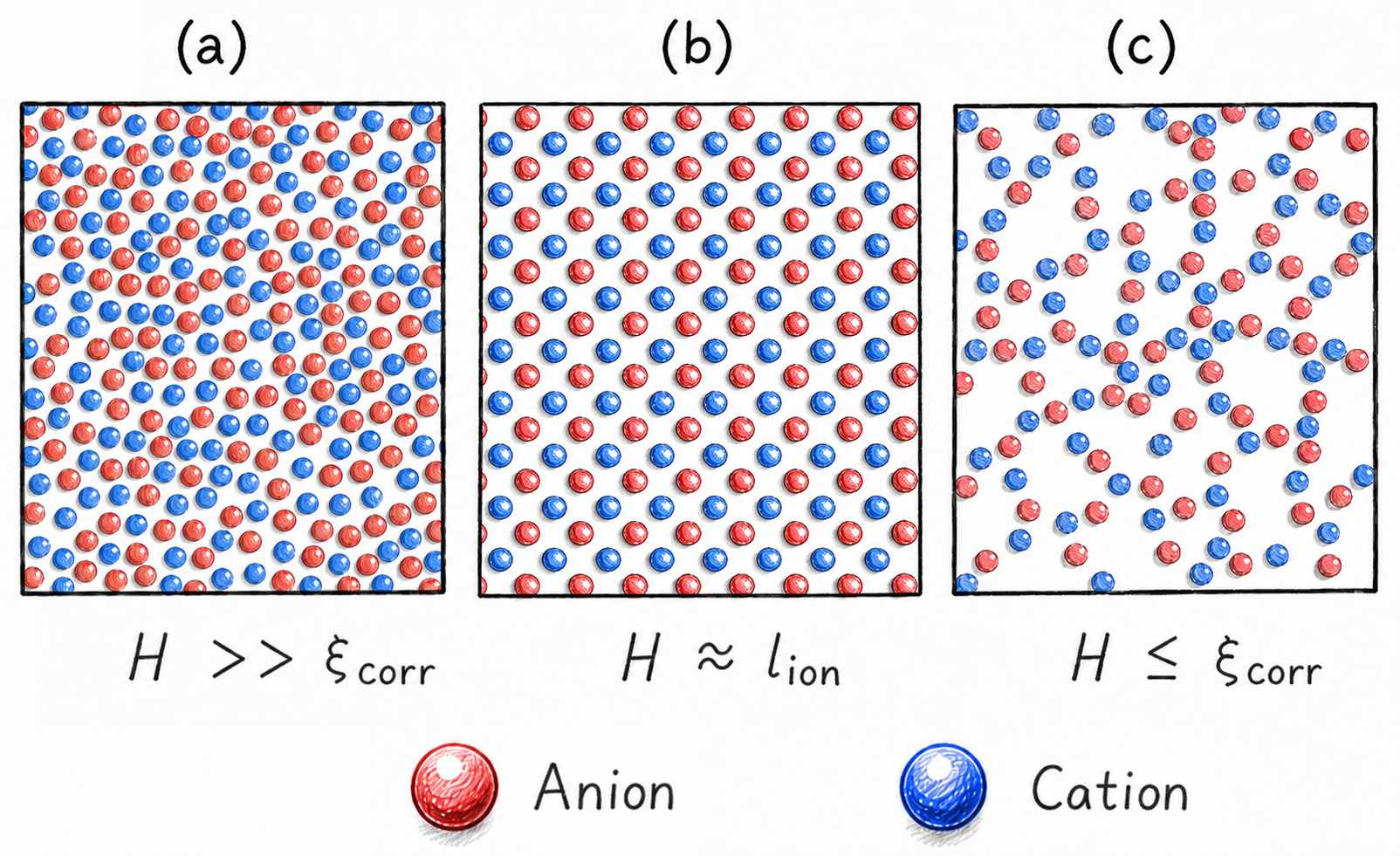}
\caption{\textbf{Confinement-induced evolution of ionic organization.} Schematic illustration of the structural organization of an electrolyte under confinement as a function of the confinement size $H$. Here, $H$ denotes the characteristic confinement size (e.g., the pore width), $\ell_{\rm ion}$ the characteristic ionic length scale (approximately the effective ion diameter), and $\xi_{\rm corr}$ the electrostatic correlation length associated with collective ionic organization, which reduces to the Debye screening length in the dilute limit. \textbf{(a) Bulk-like organization ($H \gg \xi_{\rm corr}$).} The electrolyte exhibits an isotropic, dense, and disordered structure characterized by short-range coordination and weak intermediate-range correlations. Cations (blue) and anions (red) are homogeneously distributed, forming a nanostructured liquid without long-range order. \textbf{(b) Layered organization ($H \simeq \ell_{\rm ion}$).} When the confinement size becomes comparable to the ionic diameter, steric constraints induce quasi-two-dimensional ordering (cross-sectional view perpendicular to the pore direction). The system forms densely packed layers parallel to the confining surfaces, with strong positional correlations within each layer. \textbf{(c) Correlated ionic network ($H \lesssim \xi_{\rm corr}$).} Under stronger confinement, the electrolyte evolves into a heterogeneous, percolating ionic network. Cations and anions form interconnected clusters and channels with reduced local ordering and enhanced like-charge proximity, reflecting the increasing importance of collective electrostatic correlations. The bottom axis summarizes the crossover from bulk-like organization to confinement-dominated ionic structures driven by the interplay between steric constraints and electrostatic correlations.}
\label{fig:confinement}
\end{figure*}
\section{A hierarchy of length scales}
Building on previous work on ion pairing, clustering, and mesoscale ionic organization~\cite{skarmoutsos2025,tenney2013,mceldrew2020,mceldrew2021jes,mceldrew2021jpcb}, the discussion developed so far naturally leads to a simple but fundamental conclusion: electrolyte structure cannot be understood at a single length scale. Rather, the physically relevant object evolves systematically with the observation scale, making a hierarchical description both natural and necessary. A schematic representation of this hierarchy is shown in Fig.~\ref{fig:coordination}, where the relevant objects evolve from first-shell coordination motifs to ion pairs, clusters, and finally correlated domains governing screening and collective transport.

At the smallest scale, the basic object is a \emph{motif}, i.e., the local coordination environment of Li$^+$, including solvent molecules and, at higher concentration, anions. This is governed primarily by short-range chemistry, including ionic size, donor strength, steric constraints, and solvent or polymer flexibility. In carbonate liquids it corresponds to the carbonyl-oxygen cage surrounding Li$^+$, and is typically of the order of a few \AA~\cite{skarmoutsos2015,jiang2016,ponnuchamy2018}. In polymer electrolytes, the same scale is defined by coordinating ether oxygens, nitriles, or other donor groups embedded within a slower and more constrained host matrix.

The next level is the \emph{cluster} scale. Here, local coordination motifs no longer exist as isolated entities but assemble into contact ion pairs, solvent-separated ion pairs, and larger associated ionic aggregates. At this scale, increasing salt concentration directly alters the local coordination environment. Local coordination is no longer determined solely by short-range chemistry, but also by the growing probability that neighboring ions participate in the first coordination shell. Whether an anion remains outside the first coordination shell or becomes part of it defines a genuine structural crossover, marking the transition from isolated coordination motifs to associated ionic structures. Cluster analysis in battery electrolytes has shown that even chemically simple mixtures can generate rich populations of associated ionic species~\cite{tenney2013,mceldrew2020,mceldrew2021jes,mceldrew2021jpcb}. In polymer electrolytes these associated structures may become long-lived and dominate the charge balance of the ionic sub-ensemble~\cite{coste2026,tenney2013,mceldrew2020,mceldrew2021jes,mceldrew2021jpcb}.

At still larger distances one reaches \emph{correlated domains}. These should not necessarily be interpreted as distinct thermodynamic phases; they may instead be dynamically evolving regions over which charge correlations, mesoscale order, or partial network connectivity persist. In liquid-crystalline electrolytes, for example, ordered ionic domains and lamellae explicitly introduce this scale~\cite{khan2024}. More generally, correlated domains provide the natural level at which screening and collective transport should be discussed, because it is this emergent organization---rather than isolated ions or individual coordination motifs---that governs the propagation and relaxation of charge fluctuations.

This hierarchical description is valuable because it unifies phenomena that are often treated independently. Local coordination defines the elementary charged motifs; these motifs assemble into clusters; and clusters, in turn, organize into correlated domains that govern macroscopic observables such as screening lengths, ionic conductivity, transference numbers, and the validity or breakdown of Nernst--Einstein-like descriptions. The boundaries between these levels are naturally diffuse, and the characteristic length scales depend on electrolyte composition, concentration, temperature, and morphology. The hierarchy shown in Fig.~\ref{fig:coordination} should therefore not be viewed as a collection of unrelated structural scales, but rather as a physically motivated coarse-graining sequence that progressively connects local chemistry, mesoscale organization, collective electrostatic response, and ultimately transport~\cite{skarmoutsos2025,mceldrew2020,mceldrew2021jes,mceldrew2021jpcb}.
\section{Screening: from Debye to correlated ionic domains}
\subsection{Why Debye--H\"uckel becomes insufficient}
The Debye--H\"uckel picture remains the appropriate asymptotic starting point for low ionic strength~\cite{levin2002,hansen2013}. In this limit, ions behave approximately as weakly coupled point charges embedded in a dielectric continuum, and electrostatic interactions are screened over the Debye length. Battery electrolytes, however, typically operate far from this regime. The local dielectric response is heterogeneous, ion sizes are not negligible, molecular shapes become important, and ion pairing or clustering may substantially modify the effective electrostatic environment.

Within this perspective, two closely related questions are particularly relevant. First, how should one describe the crossover from monotonic Debye-like screening to the oscillatory behavior characteristic of strongly correlated ionic fluids~\cite{rotenberg2018underscreening}? Second, what is the physical origin of the unexpectedly large screening lengths (referred to as {\em underscreening}) inferred from surface-force-balance~\cite{hayler2024surface} experiments in concentrated electrolytes and ionic liquids~\cite{Perkin2012,smith2016,perkin2017,rotenberg2018underscreening,kjellander2018,kjellander2019,safran2023}, which in several studies appear to increase with salt concentration? Although the terminology of {\em regular} and {\em anomalous} underscreening has become widely adopted, the microscopic interpretation of these observations remains actively debated~\cite{gebbie2013reply,Gebbie2013,gebbie2017long,kjellander2018,kjellander2019,safran2023}.
\subsection{A cluster-based interpretation of screening}
The simulation study of~\cite{skarmoutsos2025} provides a particularly useful framework because it directly connects structural, dielectric, and transport properties within a single model system over a broad concentration range. Several observations emerging from that work are especially relevant for the perspective developed here.

First, the electrolyte does not simply become ``more correlated'' as concentration increases. Rather, its organization evolves through identifiable structural regimes characterized by the growth, aggregation, and eventual {\em percolation} of ionic clusters~\cite{skarmoutsos2025,tenney2013,mceldrew2020,mceldrew2021jes,mceldrew2021jpcb}. Second, the characteristic length scale governing electrostatic screening can be interpreted more naturally when effective charges are assigned to correlated ionic domains rather than to bare ions. Within this picture, the relevant electrostatic objects are no longer isolated ions but dynamically evolving correlated assemblies whose effective charge and spatial extent depend on the underlying ionic organization.

Third, the largest deviations between the true conductivity and the Nernst--Einstein estimate occur precisely where electrostatic correlations are strongest and the cluster-based description becomes essential. In this regime, the effective charge carrier undergoes a progressive renormalization with increasing concentration, evolving from a locally coordinated solvated ion to a partially neutralized ion pair and ultimately to correlated ionic clusters that constitute the relevant objects for both transport and screening~\cite{skarmoutsos2025,feng2019,francelanord2019,mceldrew2020,mceldrew2021jpcb}.

This interpretation naturally connects screening with the hierarchy of correlated structures introduced in the previous section. If ion pairing and shell neutralization modify the effective charge carrier responsible for transport, they must also modify the effective object participating in electrostatic screening. In this view, screening is not an independent continuum property superimposed on the underlying chemistry, but the long-wavelength manifestation of the progressive reorganization of the electrolyte into increasingly correlated charged entities. We emphasize, however, that this interpretation should be viewed as a physically motivated framework for rationalizing the simulation results rather than as a unique explanation of all reported underscreening phenomena, whose microscopic origin remains the subject of ongoing discussion.
\begin{figure*}[t]
\centering
\includegraphics[width=0.9\textwidth]{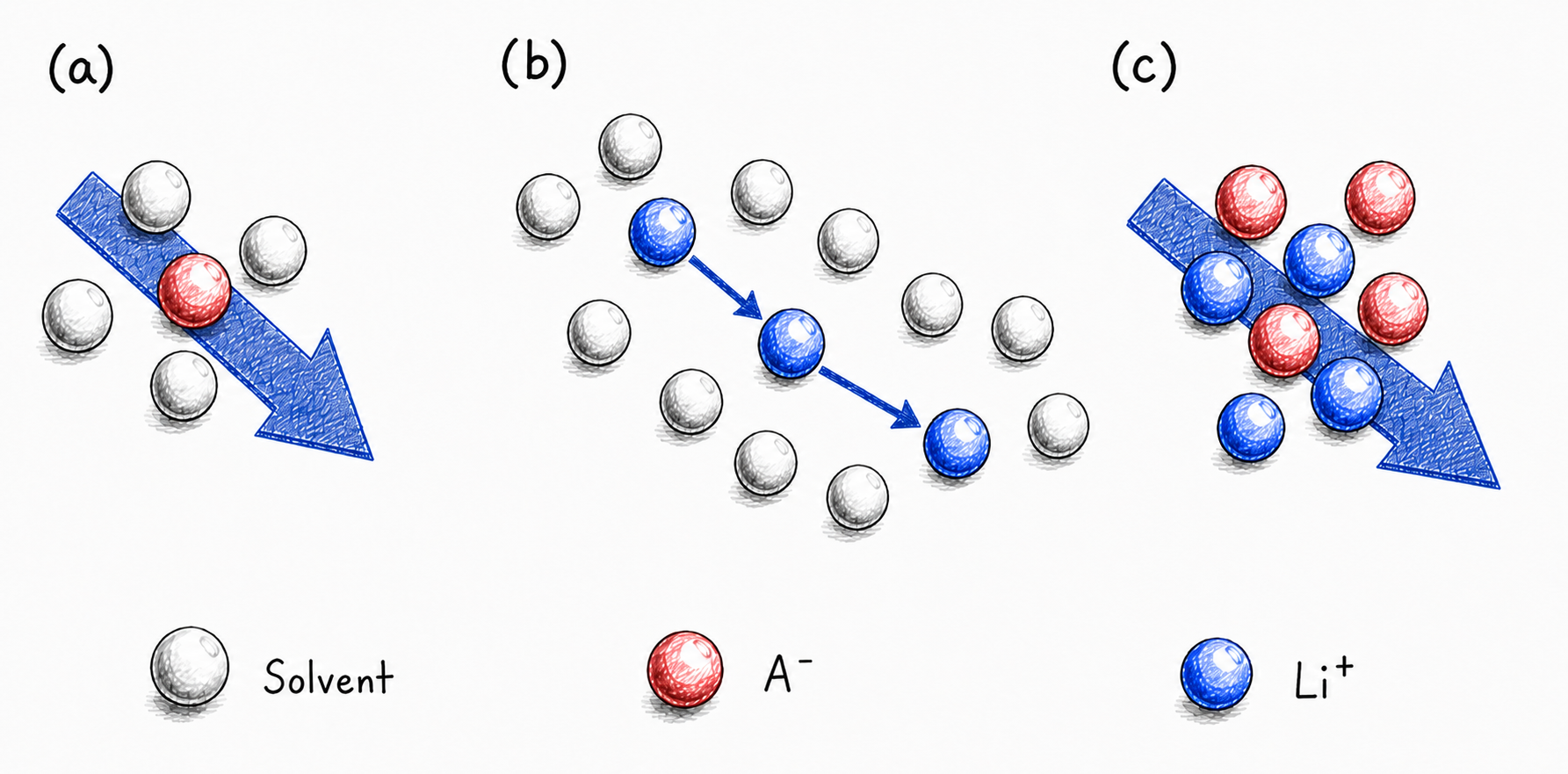}
\caption{\textbf{Schematic transport mechanisms in liquid electrolytes.} Three limiting transport mechanisms are illustrated. \textbf{(a) Vehicular transport.} Li$^+$ migrates together with its coordination shell, forming a solvated complex that moves as a single entity through the liquid. The mobility is therefore largely controlled by the diffusion of the entire coordinated species and by the viscosity of the medium. \textbf{(b) Structural exchange (hopping).} Li$^+$ migrates through successive exchanges between neighboring coordination environments. Rather than carrying an intact coordination shell, the ion transfers between transient coordination sites via a sequence of local rearrangements. \textbf{(c) Collective transport.} At high salt concentration, charge is transported by correlated ionic aggregates rather than by independent ions. Transient ion pairs, clusters, and larger correlated structures move collectively, giving rise to strong transport correlations and deviations from independent-ion behavior.}
\label{fig:transport}
\end{figure*}
\subsection{Screening as a design parameter}
From a practical standpoint, the relevance of screening extends well beyond the interpretation of surface-force-balance experiments. It governs the electrostatic coupling between charged species, influences dielectric friction, constrains ionic motion, and affects the response of electrolytes under confinement and at interfaces. Because these properties evolve with concentration and with the underlying ionic organization, electrolyte design should consider not only first-shell coordination chemistry but also the emergence and evolution of correlated ionic structures.

Several current approaches to electrolyte engineering already embody these principles. High-concentration formulations deliberately trade off some free-ion mobility for improved interfacial properties and modified solvation chemistry~\cite{cresce2012,chapman2017}. Single-ion or anchored-anion systems seek to suppress anion transport and thereby reshape the effective screening environment~\cite{ngo2024}. Liquid-crystalline electrolytes attempt to impose mesoscale order on ionic pathways~\cite{khan2024}. In each case, the objective is to engineer local coordination motifs that promote the desired collective organization. Within the framework proposed here, screening provides a physically meaningful probe of whether this structural organization has been achieved at larger length scales.
\section{Electrolytes Under Confinement}
Confinement provides a particularly stringent test of the multiscale framework developed above because it introduces an external geometric length scale that competes directly with the intrinsic ionic and correlation length scales of the electrolyte. Within our perspective, nano-confinement is therefore not a separate problem from bulk electrolyte physics, but a controlled perturbation of the same hierarchy of coordination motifs, ionic clusters, and correlated domains~\cite{skarmoutsos2025,dalmas2024,cresce2012}. At the same time, confinement may also stabilize structural organizations that have no direct bulk counterpart, making it a particularly powerful setting in which to probe the interplay between geometry and ionic correlations. Representative examples include nanoporous electrodes in supercapacitors, solid-electrolyte interphases, and nanoporous separators. 

The central competition is between the characteristic ionic length scale $\ell_{\rm ion}$, the electrostatic correlation length $\xi_{\rm corr}$, which characterizes the spatial extent of ionic correlations and reduces to the Debye length in the dilute limit, and the confinement size $H$, as illustrated in Fig.~\ref{fig:confinement}. When $H \gg \xi_{\rm corr}$, confinement has only a minor influence and the electrolyte remains essentially bulk-like. Once $H$ becomes comparable to either the ionic size or the correlation length, however, qualitatively new structural regimes emerge. Local coordination is modified by excluded-volume effects and interfacial fields, cluster populations evolve, and the effective electrostatic response may differ substantially from that of the bulk fluid.

A striking illustration is provided by ionic liquids confined in nanopores~\cite{kondrat2023theory}. The molecular dynamics simulations of~\cite{Mossa2018_PRX} showed that the interplay between ion size, pore width, and intrinsic mesoscale organization can produce re-entrant phase behavior, including capillary crystallization and ordered quasi-two-dimensional states. The resulting structure depends sensitively on whether the pore accommodates frustrated ionic packing, a layered arrangement, or a bulk-like liquid organization.

Confinement can also modify electrostatic screening in a qualitatively different manner. In metallic nanopores, image-charge effects reduce the effective electrostatic interactions between ions and may allow ions of identical charge to occupy the same pore region, giving rise to the so-called {\em superionic state} discussed in the theoretical and experimental literature~\cite{kondrat2011,kondrat2011superionic,futamura2017partial}. In this regime, the screening response is not simply the bulk response truncated by boundaries; rather, it emerges from the interplay between electrostatics, geometry, and confinement-induced ionic organization.

The broader lesson is that confinement makes the hierarchy of length scales experimentally unavoidable. Once the external dimension $H$ becomes comparable to either $\ell_{\rm ion}$ or the relevant correlation length, local coordination, mesoscale organization, screening, and ultimately transport can no longer be considered independently. Although many examples originate from ionic liquids and supercapacitors, the same physical principles apply to lithium electrolytes near interfaces and within confined morphologies: confinement modifies the same correlated structures that govern transport and screening in the bulk~\cite{kondrat2023theory,Mossa2018_PRX,kondrat2011,kondrat2011superionic,futamura2017partial,cresce2012}. Rather than introducing a new hierarchy of structures, confinement perturbs and reshapes the hierarchy already present in the bulk, thereby providing a particularly sensitive probe of the coupling between local coordination, collective electrostatic response, and ionic transport.
\section{A unified picture of lithium transport}
Within the framework developed above, transport is not an independent phenomenon superimposed on structure and screening. Rather, it is the dynamical manifestation of the same correlated objects that define local coordination, ion pairing, cluster formation, and electrostatic response. As the physically relevant object evolves from a solvated ion to larger correlated ionic assemblies, the dominant transport mechanism evolves accordingly. Figure~\ref{fig:transport} summarizes three limiting transport mechanisms that naturally emerge along this hierarchy.
\subsection{Vehicular motion}
The first limiting mechanism is vehicular transport: Li$^+$ moves together with a relatively persistent coordination shell. This picture is most natural in dilute carbonate electrolytes, where the cation behaves as a dressed particle defined by its local solvation environment~\cite{skarmoutsos2015}. It remains useful whenever solvent exchange is slow compared with translational motion, or whenever the coordination shell remains sufficiently long-lived for the cation to carry its local environment over appreciable distances.

Vehicular motion is nevertheless not equivalent to ideal independent-ion diffusion. The mobility depends on shell size, solvent viscosity, dielectric friction, and the dipolar structure of the coordination environment. Preferential solvation therefore affects not only thermodynamics but also dynamics: two coordination shells with similar mean coordination numbers may exhibit markedly different mobilities because they differ in symmetry, dipole cancellation, or solvent friction.
\begin{figure*}[t]
\centering
\includegraphics[width=0.95\textwidth]{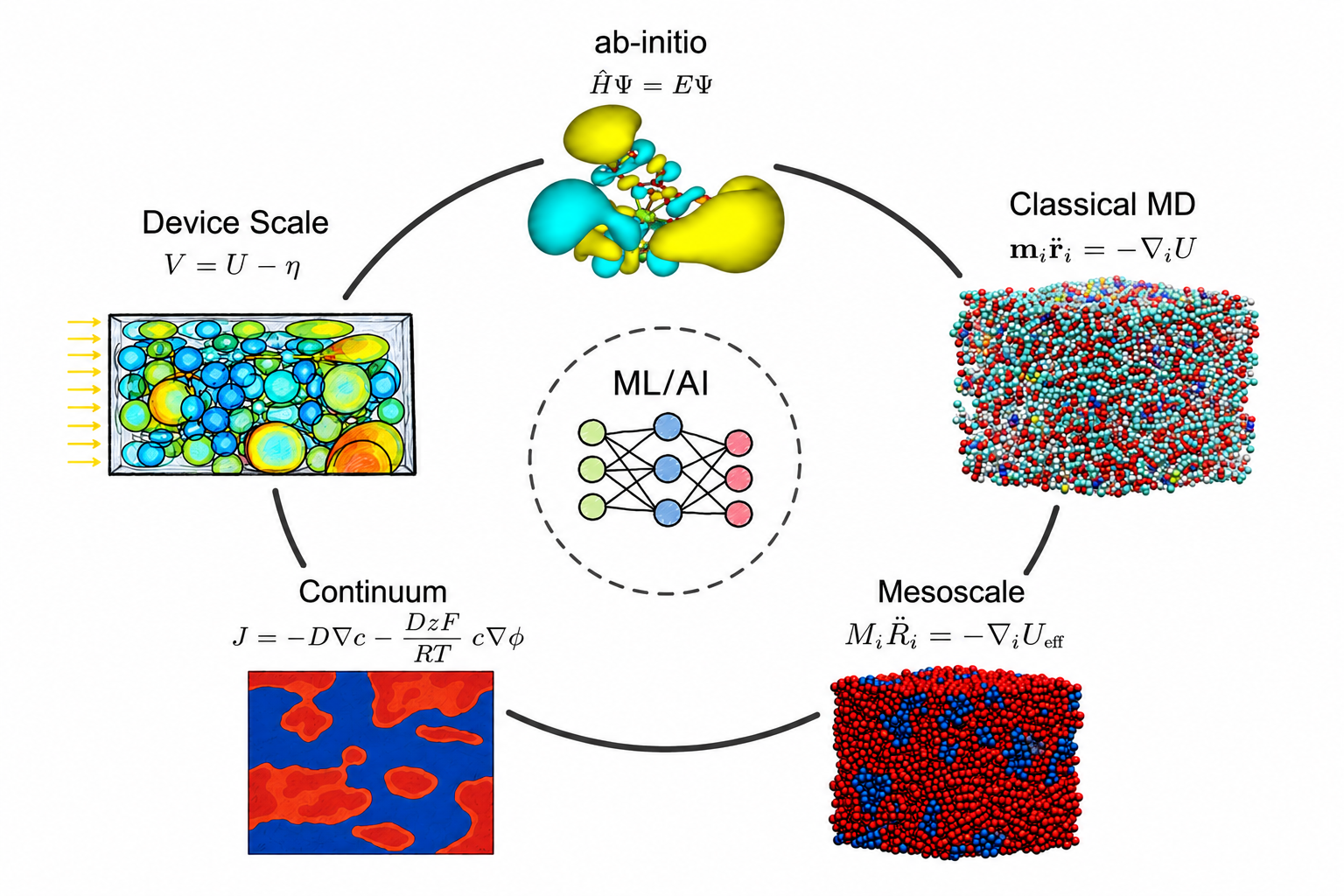}
\caption{\textbf{Machine-learning-assisted multiscale computational framework for electrolyte modeling and design.} Schematic representation of the hierarchy of computational descriptions used to investigate electrolyte systems across increasing length and time scales. Electronic-structure calculations ($\hat{H}\Psi=E\Psi$) provide the microscopic description of coordination chemistry, polarization, and reaction energetics. Classical molecular dynamics ($m_i\ddot{\mathbf r}_i=-\nabla_iU$) resolves the statistical organization, dynamics, and transport of atomistic systems. Coarse-grained models ($M_i\ddot{\mathbf R}_i=-\nabla_iU_{\rm eff}$) describe the emergence of correlated ionic structures at mesoscopic scales. Continuum transport models ($\mathbf J=-D\nabla c-(D_zF/RT)c\nabla\phi$) capture macroscopic concentration, potential, and flux fields, while device-level models ($V=U-\eta$) predict the electrochemical response of complete cells. Machine learning and artificial intelligence (ML/AI), shown at the center, provide a unifying framework for constructing interatomic potentials, discovering coarse-grained representations, learning surrogate models, and transferring physically relevant information across modeling scales. The circular layout emphasizes the bidirectional flow of information between neighboring levels, enabling predictive multiscale modeling and digital-twin approaches for next-generation electrolyte materials and electrochemical devices.}
\label{fig:computational}
\end{figure*}
\subsection{Structural exchange and hopping}
The second limiting mechanism is structural diffusion, in which Li$^+$ transport proceeds through exchanges between coordination environments rather than through the persistent motion of a single intact shell. This mechanism is especially important in polymer electrolytes, where the cation migrates between transient coordination sites defined by polymer segments~\cite{berthier1983,borodin2006,coste2026}, although analogous exchange processes also occur in structured liquids and liquid-crystalline electrolytes~\cite{khan2024}.

In this regime, transport is governed by the competition between coordination-site residence times and exchange rates. If coordination is too strong, Li$^+$ becomes kinetically trapped; if it is too weak, dissociation increases and structural connectivity is lost. Efficient transport therefore typically emerges at intermediate coordination strength, where local binding remains sufficient to maintain a connected ionic environment while still allowing rapid structural rearrangements between neighboring coordination motifs.
\subsection{Collective cluster motion}
The third limiting mechanism is collective transport. In concentrated liquids and polymer electrolytes, charge may be carried not only by nominally individual ions but also by associated ionic entities or correlated domains. This generally reduces the validity of independent-particle transport descriptions and constitutes one of the principal origins of large deviations from the Nernst--Einstein relation~\cite{francelanord2019,feng2019,coste2026,skarmoutsos2025}. Depending on cluster charge, lifetime, and connectivity, collective motion may either reduce the net conductivity through anticorrelated cation--anion motion or facilitate transport through connected ionic morphologies.

This mechanism is the natural dynamical counterpart of the cluster-based interpretation of screening developed in the previous section. If correlated domains constitute the physically relevant objects governing static electrostatic correlations, they should also dominate current relaxation, ionic conductivity, and transference. Within this perspective, the apparent failure of the Nernst--Einstein relation is therefore not merely the breakdown of an approximate formula, but rather an indication that the physically relevant transport carriers are no longer individual ions.

The same considerations apply to the lithium transference number, which cannot be interpreted as an intrinsic single-ion property in correlated electrolytes. As ion pairing, clustering, and collective dynamics develop, the effective charge carriers become correlated entities rather than individual ions~\cite{francelanord2019,feng2019,pesko2017}. The measured transference number then reflects the net contribution of these collective transport modes, including anticorrelated cation--anion motion, and may become small or even negative despite significant lithium mobility~\cite{pesko2017,feng2019}. Rather than representing an intrinsic property of an isolated lithium ion, the transference number therefore provides information on the collective transport modes sustained by the correlated electrolyte. Within this framework, the transference number becomes a direct dynamical probe of the correlated structures governing both screening and transport.
\subsection{One framework, three weights}
Vehicular, hopping, and collective mechanisms are therefore best viewed not as separate theories, but as three continuously varying weights within a common transport framework. Their relative importance depends on solvent composition, polymer relaxation, salt identity, concentration, and, more generally, on the degree of ionic correlation. Dilute liquids are usually dominated by vehicular transport; intermediate concentrations show increasing shell exchange and ion pairing; highly concentrated liquids and many polymer systems require explicit descriptions based on ionic clusters and correlated networks. Transport therefore evolves continuously together with the same correlated structures that govern screening and mesoscale organization.

A key implication of this framework is that transport does not proceed through abrupt changes of mechanism, but through a continuous redistribution of these weights governed by the stability and dynamics of correlated ionic structures. In particular, optimal transport often emerges when coordination is sufficiently robust to maintain structural connectivity while remaining sufficiently labile to permit rapid reorganization. At higher concentration, where correlations extend over multiple length scales, transport becomes increasingly sensitive to the topology, connectivity, and lifetime of ionic aggregates and correlated domains, and single-particle descriptions progressively lose predictive power. In this regime, quantities such as the lithium transference number and the notion of single-ion transport must be interpreted in terms of correlated transport carriers rather than individual ions.

From a design perspective, this picture implies that improving electrolyte performance requires controlling not only local coordination chemistry but also the mesoscale organization and collective dynamics that emerge from it. In this sense, transport becomes the dynamical signature of the same hierarchy of correlated structures that governs coordination and screening. We return to this idea in more concrete terms in the following section.
\section{A Multiscale Computational Framework for Correlated Electrolytes}
The hierarchy of length scales discussed above provides a unified picture of electrolyte structure. A natural next step is to translate this picture into a predictive computational framework~\cite{Franco2019}, depicted in Fig.~\ref{fig:computational}, capable of \emph{propagating information} from the nanoscale, where chemical specificity resides, to the macroscale, where transport, screening, and electrochemical performance are experimentally probed. Figure~\ref{fig:computational} emphasizes that each level of description is governed by a different mathematical model, ranging from the Schr\"odinger equation at the electronic-structure level to continuum transport equations and device-scale models. The central challenge is therefore not simply to bridge spatial and temporal scales, but to identify the physically relevant variables and representations that should be transferred consistently between fundamentally different descriptions of the same electrolyte.
\paragraph*{{\bf Electronic-structure level: defining coordination motifs.}} Electronic-structure calculations provide the most fundamental description of the electrolyte by resolving ion--solvent and ion--ion interactions at quantum-mechanical scales~\cite{borodin2009,borodin2016,ganesh2011}. Beyond binding energies, they provide information on polarization response, charge distribution, electronic structure, and the free-energy landscape associated with different coordination environments. Within the present framework, these calculations define the elementary coordination motifs that constitute the microscopic building blocks of the electrolyte. Importantly, the relevant information extends well beyond isolated clusters: local dipoles, anisotropy, and environmental response must also be characterized, since they influence the assembly of correlated structures at larger scales.

Recent theoretical developments further illustrate this broader role by quantitatively connecting local solvation structure, ionic association, thermodynamic activities, and electrochemical stability through a combined framework based on electronic-structure calculations, molecular dynamics, statistical thermodynamics, and spectroscopy~\cite{PRXEnergy.2.013007,Goodwin2026}. Electronic-structure methods therefore provide not only local energetics, but also the microscopic information required to predict macroscopic electrochemical behavior.
\paragraph*{{\bf Atomistic simulations: statistical realization of motifs and clusters.}} Classical molecular dynamics simulations extend this description by sampling the statistical ensemble of coordination environments, cluster populations, and dynamical processes~\cite{skarmoutsos2015,skarmoutsos2025}. This level provides direct access to coordination-number distributions, ion-pairing statistics, cluster lifetimes, dielectric response, and current correlations entering transport coefficients. At this scale, coordination motifs become dynamically interacting entities that assemble into clusters and, at higher concentration, extended correlated domains. Deviations from Nernst--Einstein behavior and the emergence of collective transport can then be directly related to these correlations. Despite their richness, atomistic simulations remain limited in accessible time and length scales and do not directly provide a reduced description of the effective carriers governing macroscopic behavior.
\paragraph*{{\bf Coarse-grained representations and emergent carriers.}} A central step in the multiscale framework is the definition of coarse-grained variables that capture intermediate- and large-scale behavior. Within the present perspective, these variables naturally correspond to ionic clusters and correlated domains governing screening and transport. This requires identifying physically meaningful structural objects through cluster analysis or connectivity criteria~\cite{tenney2013,mceldrew2020,mceldrew2021jes}, and assigning them effective properties such as charge, mobility, and interaction potentials. This mapping is generally not unique: effective charge depends on internal correlations and polarization, while dynamical properties depend on both internal structure and coupling to the surrounding medium. Preserving the physical content of this coarse-graining is therefore essential.
\paragraph*{{\bf Machine learning as a unifying layer.}} The central challenge of multiscale electrolyte modeling is not simply the generation of larger simulation datasets, but the identification and transfer of the physically relevant information between successive levels of description. Within this framework, machine learning (ML) and artificial intelligence (AI) therefore play a fundamentally different role from that of an additional simulation technique~\cite{Lombardo2021}. Rather, they provide a systematic framework for discovering, compressing, and propagating the representations that connect electronic-structure calculations, atomistic simulations, coarse-grained models, continuum descriptions, and ultimately device-scale simulations.

At the atomistic level, machine-learned interatomic potentials enable simulations that retain near \emph{ab-initio} accuracy while reaching system sizes and time scales inaccessible to direct electronic-structure methods~\cite{behler2007,noe2020}. Importantly, the descriptors used by these models are not merely intermediate features for predicting energies and forces, but also provide compact representations of local atomic environments that can be mined to reveal structural motifs and reaction pathways, effectively bridging atomistic simulations and data-driven coarse-graining~\cite{warnicka2026following}. 

More fundamentally, representation-learning approaches can identify collective variables describing coordination environments, ionic clusters, correlated domains, and transport pathways directly from simulation data, thereby providing physically meaningful coarse-grained variables instead of prescribing them \emph{a priori}. Graph-based representations constitute one particularly natural realization of this philosophy, since atoms, molecules, coordination motifs, ionic clusters, and correlated domains can all be represented within a common mathematical framework~\cite{husic2020,bapst2020,xie2021}.

As illustrated schematically in Fig.~\ref{fig:computational}, ML/AI occupies the interfaces between neighboring modeling levels rather than constituting an additional modeling level itself. In this sense, its primary role is to enable the consistent transfer of physically relevant information across scales while preserving the descriptors that govern electrolyte behavior.
\paragraph*{{\bf Interfaces between scales: consistency and information transfer.}} A critical aspect of any multiscale framework is the treatment of couplings between different levels of description. These couplings are best viewed as mappings between distinct mathematical representations of the same physical system rather than as independent models. The objective is not simply to exchange parameters between models, but to preserve the physical information encoded in the correlated ionic structures identified at finer levels of description.

Three mappings are particularly important. First, electronic-structure calculations provide information on local energetics, polarization, and charge transfer that must be translated into classical force fields or machine-learned interatomic potentials for large-scale simulations~\cite{behler2016,bartok2017review}. Second, atomistic simulations must be connected to coarse-grained descriptions in which molecular configurations are projected onto clusters or correlated domains~\cite{voth2008,wang2019}. Third, coarse-grained models must be linked to continuum descriptions, where transport, screening, and electrochemical response are defined at experimentally relevant scales.

Consistency across these mappings requires preserving key observables under coarse-graining, including thermodynamic quantities, structural correlations, dielectric response, and transport coefficients. Reduced models should therefore reproduce not only static structure but also dynamical observables such as current autocorrelation functions, collective diffusion coefficients, and dielectric relaxation. ML provides a natural framework for learning such mappings while simultaneously constraining multiple observables~\cite{noe2020}.
\paragraph*{{\bf From correlated structures to macroscopic response.}} Within this framework, information propagates through a sequence of transformations linking microscopic structure to macroscopic response. Local coordination determines elementary charged species, cluster formation modifies their effective charge and mobility, and mesoscale correlations shape collective transport. 

These mesoscale descriptions naturally connect to continuum approaches such as phase-field models, where coarse-grained order parameters encode the collective evolution of concentration, phase composition, or interfacial structure while preserving the essential physics inherited from atomistic descriptions. More generally, multiscale frameworks based on systematic information transfer between electronic-structure calculations, atomistic simulations, mesoscale models, and continuum descriptions provide a consistent route toward predictive materials modeling across length and time scales~\cite{Yang2026}.

Quantities such as screening length, conductivity, transference number, thermodynamic activity, and electrochemical stability therefore emerge from this sequence rather than constituting independent inputs. The correlation length associated with ionic domains naturally extends the Debye length beyond the dilute limit, while deviations from Nernst--Einstein behavior reflect the growing importance of correlated transport.

Several challenges remain in realizing this framework quantitatively. A first concerns the accurate treatment of polarization and many-body effects across scales, which remain essential for predictive electrolyte modeling~\cite{Madden2014}. A second is the identification of transferable coarse-grained variables that remain robust across chemistries and thermodynamic conditions~\cite{Noid2013}. A third challenge is the treatment of rare events and long-time dynamics, which are often central to transport yet difficult to sample directly~\cite{Tiwary2015,Devergne2026}.

Despite these challenges, the combination of electronic-structure methods, atomistic simulations, coarse-grained modeling, and machine learning offers a promising route toward predictive multiscale modeling of electrolytes~\cite{Butler2018}. Within our perspective, the hierarchy of correlated ionic structures naturally defines the hierarchy of computational representations, while ML/AI provides the machinery required to discover, preserve, and transfer the physically relevant information between successive levels of coarse graining. This philosophy naturally extends to digital-twin frameworks~\cite{Lombardo2021}, in which continuously updated multiscale models integrate simulation, experiment, and data-driven learning to accelerate the rational design of next-generation electrolyte materials and electrochemical devices.
\section{What this implies for electrolyte design}
The multiscale framework developed throughout this perspective naturally leads to a different way of thinking about electrolyte design. Rather than optimizing isolated molecular properties, the objective is to control the coupled evolution of local coordination, ionic association, mesoscale organization, and collective transport. Within this framework, molecular chemistry defines the elementary coordination motifs, but electrolyte performance ultimately emerges from the hierarchy of correlated structures that these motifs generate. The following design principles therefore represent successive levels of the same hierarchy rather than independent optimization strategies.
\paragraph*{{\bf 1.~Design coordination environments rather than coordination numbers.}} Four-fold coordination provides a useful reference in dilute carbonate electrolytes, but under technologically relevant conditions the more informative descriptor is the complete distribution of solvent--anion coordination motifs and donor environments surrounding Li$^+$~\cite{jiang2016,ponnuchamy2018,kameda2007,cresce2015anion,chapman2017}. Effective electrolyte design should therefore target the statistics of coordination motifs rather than a single average coordination number.
\paragraph*{{\bf 2.~Optimize local electrostatics as well as local composition.}} Mixed-carbonate electrolytes demonstrate that transport is governed not only by which molecules coordinate Li$^+$, but also by the geometry and electrostatic organization of the coordination shell. Shell dipole, local symmetry, molecular packing, and dipolar cancellation all influence mobility and should therefore be regarded as design variables alongside chemical composition~\cite{borodin2009,borodin2016,skarmoutsos2015}.
\paragraph*{{\bf 3.~Control ionic association rather than simply suppressing ion pairing.}} The entry of anions into the first coordination shell represents a structural reorganization rather than merely a chemical equilibrium. Once ion pairing develops, the effective charge carrier changes its identity, directly affecting conductivity, lithium transference, screening, and interfacial behavior~\cite{jiang2016,ponnuchamy2018,chapman2017}. The objective is therefore not necessarily to eliminate ion pairing, but to control its extent, lifetime, connectivity, and resulting transport properties.
\paragraph*{{\bf 4.~Use correlated structures as transport descriptors.}} Large deviations from the Nernst--Einstein relation often indicate that the physically relevant transport units are correlated ionic clusters or domains rather than nominally independent ions~\cite{francelanord2019,feng2019,mceldrew2020,mceldrew2021jpcb,skarmoutsos2025}. Cluster populations, connectivity, effective charge, and correlation lengths should therefore be regarded as essential descriptors for electrolyte optimization.
\paragraph*{{\bf 5.~Engineer mesoscale organization.}} Liquid-crystalline electrolytes, hybrid polymer--ceramic systems, and confined electrolytes demonstrate that mesoscale morphology can either facilitate or hinder transport depending on how it reorganizes local coordination and ionic correlations~\cite{khan2024,dalmas2024}. Electrolyte architectures should therefore be designed to promote favorable correlated ionic pathways rather than treating morphology as a secondary consequence of chemistry.

Taken together, these principles define a hierarchical design strategy. Local chemistry determines coordination motifs; coordination motifs determine ionic association; ionic association governs mesoscale organization; and mesoscale organization ultimately controls screening, transport, interfacial response, and electrochemical performance. Rational electrolyte design should therefore focus not on optimizing a single molecular descriptor, but on engineering the entire hierarchy of correlated structures responsible for macroscopic behavior.
\section{Open questions and broader applicability across electrolyte systems}
Although the multiscale framework proposed here provides a coherent physical picture of electrolyte organization, several important questions remain open. Most of them concern the extent to which the hierarchy of correlated structures developed throughout this Perspective remains predictive across different chemistries, thermodynamic conditions, and electrochemical environments.

A first important question concerns the nature of electrostatic screening in concentrated electrolytes, and in particular the relation between screening lengths inferred from surface-sensitive measurements and those extracted from bulk structural and dielectric correlations. The cluster-based interpretation proposed in~\cite{skarmoutsos2025} provides one physically consistent framework, but its validity remains to be assessed across a broader range of electrolyte chemistries, concentrations, temperatures, and confinement conditions. Related cluster and network approaches point in a similar direction~\cite{mceldrew2020,mceldrew2021jes,mceldrew2021jpcb}, suggesting that correlated ionic structures may provide a common physical language for understanding screening beyond the Debye--H\"uckel regime.

A second important question concerns the coupling between coordination dynamics, electrochemical stability, interfacial properties, ion transport across electrolyte--electrode interfaces, and lithium insertion/extraction processes. The same coordination motifs that govern bulk transport also determine the local chemical environment, interfacial ion-transfer kinetics, desolvation barriers, and the structures from which the solid-electrolyte interphase and other interphases emerge~\cite{cresce2012,Xu2014}. 

Recent theoretical developments further indicate that local solvation structure can be quantitatively connected to thermodynamic activities and electrochemical stability through a combination of electronic-structure calculations, molecular simulations, and statistical thermodynamics~\cite{PRXEnergy.2.013007,Goodwin2026}. A predictive theory of electrolyte performance should therefore connect coordination statistics, ionic correlations, screening, thermodynamics, interfacial transport, lithium insertion/extraction kinetics, and decomposition pathways within a unified framework, rather than treating them as separate problems.

A third open question is computational. Electronic-structure calculations remain indispensable for describing donor preferences, polarization, and local coordination motifs, whereas classical molecular dynamics simulations are required to resolve mesoscale organization, dielectric response, and collective transport. Developing transferable coarse-grained representations that faithfully preserve this information across scales, particularly in the presence of polarization and many-body interactions, remains one of the central challenges for predictive electrolyte modeling~\cite{ganesh2011,borodin2016,bhatt2015}. As discussed in the previous section, machine learning provides a particularly promising route toward discovering these multiscale representations directly from simulation and experimental data while ensuring consistent information transfer between different levels of description.

A fourth question concerns the generality of the framework itself. Although the present Perspective has focused primarily on lithium-based electrolytes, the underlying concepts are expected to apply much more broadly across alkali-metal, multivalent, and ionic-liquid electrolytes~\cite{Xu2004,Xu2014}. The quantitative balance between coordination, ionic association, and collective organization will naturally depend on ion size, valence, charge density, solvent chemistry, and dielectric environment, but the hierarchical organization of correlated structures is expected to remain a useful organizing principle.

Sodium-ion electrolytes provide an instructive comparison. Owing to the larger ionic radius and lower charge density of Na$^{+}$, coordination environments are generally more weakly bound and dynamically more labile than in lithium systems. Consequently, sodium electrolytes often exhibit larger coordination numbers, weaker ion pairing, and a different balance between vehicular, structural, and collective transport mechanisms~\cite{Ponrouch2015,Delmas2018}. Potassium electrolytes are expected to follow similar trends, with their even larger ionic radius further favoring solvent-separated ionic structures.

Multivalent electrolytes introduce additional complexity. In magnesium- and zinc-based electrolytes, stronger Coulomb interactions produce more tightly bound coordination shells, enhanced ion pairing, and stronger electrostatic correlations~\cite{Canepa2017}. As a consequence, cluster formation and correlated ion motion may become important already at relatively modest salt concentrations, potentially leading to substantial deviations from dilute-solution transport models. Understanding how coordination chemistry, polarization, and collective electrostatics combine to determine transport and electrochemical performance in these systems remains a major challenge for next-generation battery technologies.

Finally, the perspective developed here suggests that materials discovery itself may require a different generation of descriptors. If electrolyte performance is governed not simply by dielectric constant, donor number, or isolated binding energies, but by distributions of coordination motifs, shell dipoles, cluster charge statistics, correlation lengths, and collective transport modes, then these quantities should become the natural descriptors for both physics-based electrolyte design and data-driven materials discovery. Identifying, predicting, and optimizing such descriptors represents a promising direction for future multiscale modeling, machine-learning-assisted coarse graining, and AI-driven electrolyte design.
\section{Conclusions}
The traditional separation between coordination chemistry, screening theory, transport theory, and computational modeling is becoming increasingly inadequate for describing modern electrolytes. Throughout this Perspective, we have argued that these subjects are more naturally understood within a common multiscale framework, in which local coordination motifs, ion pairs, ionic clusters, correlated domains, and collective transport represent successive levels of organization rather than independent physical phenomena. Within this hierarchy, screening, transport, and ultimately electrochemical performance emerge not from isolated bare ions, but from the statistics, dynamics, and collective organization of correlated ionic structures.

The physical picture may be summarized simply. At low concentration, Li$^+$ is reasonably described as a cation dressed by a local coordination shell. As concentration increases, anion participation progressively modifies that shell and changes the identity of the effective charge carrier. These coordination motifs assemble into ion pairs, ionic clusters, and, in some systems, extended correlated domains. As the characteristic structural scale increases, so too does the physically relevant transport object, leading naturally to the emergence of collective screening, correlated transport, and significant departures from dilute-solution descriptions.

The practical implications are equally straightforward. Rational electrolyte design should no longer focus exclusively on optimizing isolated molecular descriptors such as donor number, dielectric constant, or average coordination number. Instead, it should seek to control the complete hierarchy of correlated structures, from local coordination chemistry to mesoscale ionic organization and collective dynamics. Within this Perspective, coordination chemistry, screening, transport, thermodynamics, and interfacial behavior become different manifestations of the same underlying structural organization.

The same hierarchy naturally defines a multiscale computational strategy. Electronic-structure calculations identify elementary coordination motifs; atomistic simulations determine their statistical organization and dynamics; coarse-grained models describe the emergent correlated carriers; and continuum theories capture macroscopic electrochemical behavior. Machine learning and artificial intelligence provide a natural framework for identifying, preserving, and transferring the physically relevant information between these successive levels of description, thereby enabling predictive multiscale models and, ultimately, digital twins of electrolyte materials and electrochemical devices.

Many important challenges remain, including the quantitative description of correlated screening, the coupling between local coordination and electrochemical stability, the development of transferable coarse-grained representations, and the identification of robust descriptors for data-driven materials discovery. Nevertheless, the framework proposed here suggests that these problems should not be addressed independently, but as complementary manifestations of the same multiscale organization of correlated charged fluids.

More broadly, the perspective developed here advocates a shift in how electrolytes are conceptualized. Rather than viewing them as collections of independently solvated ions, we propose that they be regarded as hierarchically organized correlated systems whose physically relevant building blocks evolve continuously across length and time scales. We hope that this viewpoint will contribute to a more unified understanding of electrolyte physics while providing a coherent foundation for predictive multiscale modeling, machine-learning-assisted materials discovery, and the rational design of next-generation electrolytes across a broad range of electrochemical technologies.
\begin{acknowledgments}
This work was supported by the French National Research Agency under the France 2030 program (Grant ANR-22-PEBA-0002), and by the LabEx MateriAlps 2025–2026 program.
\end{acknowledgments}
\bibliography{perspective_refs}
\end{document}